\documentclass[11pt]{article}
\usepackage[a4paper,top=3cm,bottom=3cm,left=3cm,right=3cm,marginparwidth=1.75cm]{geometry}
\usepackage{graphicx} 
\usepackage{amsmath}
\usepackage{hyperref}
\hypersetup{
    colorlinks=true,
    linkcolor=blue,
    filecolor=blue,      
    urlcolor=blue,
    allcolors=blue,
    pdftitle={Overleaf Example},
    pdfpagemode=FullScreen,
    }

\usepackage{apacite}
\usepackage{authblk}
\usepackage{lineno}

\usepackage{caption}
\title{\textbf{Shallow Cumulus Cloud Fields Are Optically Thicker When They Are More Clustered}}

\author[1]{Pouriya Alinaghi\footnote{Correspondence: p.alinaghi@tudelft.nl}}
\author[1,2]{Martin Janssens}
\author[3]{Goutam Choudhury}
\author[3]{Tom Goren}
\author[1,4]{A. Pier Siebesma}
\author[1]{Franziska Glassmeier}

{\small
\affil[1]{Delft University of Technology, Delft, The Netherlands}
\affil[2]{Wageningen University \& Research, The Netherlands}
\affil[3]{Bar-Ilan University, Ramat Gan, Israel}
\affil[4]{Royal Netherlands Meteorological Institute (KNMI), De Bilt, The Netherlands}
}

\date{Date: \today}

\begin{document}

\maketitle
\emph{Note: In the final stages of preparing this manuscript, we became aware of independent, related work by \citeA{denby2023charting} (\url{https://doi.org/10.48550/arXiv.2309.08567}).}

\newpage
\section*{abstract}
Shallow trade cumuli over subtropical oceans are a persistent source of uncertainty in climate projections. Mesoscale organization of trade cumulus clouds has been shown to influence their cloud radiative effect (CRE) through cloud cover. We investigate whether organization can explain CRE variability independently of cloud cover variability. By analyzing satellite observations and high-resolution simulations, we show that increased clustering leads to geometrically thicker clouds with larger domain-averaged liquid water paths, smaller cloud droplets, and consequently, larger cloud optical depths. The relationships between these variables are shaped by the mixture of deep cloud cores and shallower interstitial clouds or anvils that characterize cloud organization. Eliminating cloud cover effects, more clustered clouds reflect up to 20 W/m$^2$ more instantaneous shortwave radiation back to space.

\section{Introduction}

Marine shallow cumulus clouds, as the most prevalent cloud type \cite{johnson1999trimodal}, play a vital role in the climate system by reflecting incoming solar radiation back to space \cite{bony2004dynamic,bony2005marine,bony2015clouds}. The response of these clouds to changes in cloud-controlling factors remains as one of the largest sources of uncertainty of climate projections \cite{schneider2017climate,nuijens2019boundary,sherwood2020assessment}, despite recent advances (e.g.,\citeauthor{vogel2022strong}, 2022). 

Shallow cloud fields in the trades exhibit a diverse range of patterns, which have subjectively, and almost poetically, been classified as \textit{Sugar}, \textit{Gravel}, \textit{Flowers} and \textit{Fish} \cite{stevens2020sugar}. A comprehensive analysis by \citeA{janssens2021cloud} shows that the quantification of such patterns  needs at least two effective dimensions. Cloud fraction $f_c$ as a bulk 1D measure and the organization index $I_{org}$, which quantifies the level of non-randomness in cloud spatial distribution within a cloud field \cite{weger1992clustering,tompkins2017organization} are an example of a suitable variable choice to represent these two dimensions. How relevant this mesoscale organization is for the low-cloud climate feedback remains an open question.

The shortwave (SW) and longwave (LW) radiative effect of clouds is sensitive to organization \cite{denby2020discovering}. The daily mean cloud radiative effect (CRE) varies by approximately 10 W/m$^2$ among subjectively classified cloud patterns, primarily due to differences in $f_c$, with a variability of 5 to 10 W/m$^2$ at a fixed $f_c$ \cite{bony2020sugar}. Contrary to the case of deep convective clouds \cite{tobin2012observational}, where outgoing LW radiation increases with clustering, \citeA{luebke2022assessment} suggest a correlation between increased $I_{org}$ values and \textit{reduced} LW warming. No influence of clustering on SW cooling is observed in their study. For stratocumulus cloud decks, \citeA{mccoy2022role} demonstrate that different morphologies, indicative of differences in the horizontal organization of the cloud decks, modulate the relationship between albedo and $f_c$.

We aim to investigate whether -- independent of $f_c$ variability -- the horizontal organization of shallow cumulus cloud fields has an impact on the their net CRE. To do so, we combine satellite data with a large ensemble of large-eddy simulations (section \ref{sec:data}). After removing the confounding effect of cloud fraction (section \ref{sec:cre-org}), we show that clustered cloud fields feature optically thicker clouds (section \ref{sec:vert-org}). This stems from clustered cloud fields containing more liquid-water path and smaller retrieved cloud droplets (section \ref{sec:vert-org}). Analyzing the simulations establishes that increased liquid-water path due to increased clustering primarily results from increased cloud geometric thickness (section \ref{sec:geo-org}). Section~\ref{Conclusion} concludes.
\section{Description of the Data}
\label{sec:data}

Following previous studies \cite{stevens2020sugar,bony2020sugar,janssens2021cloud}, we focus on clouds over the tropical Atlantic Ocean to the east of Barbados (10$^{\circ}$-20$^{\circ}$N, 48$^{\circ}$-58$^{\circ}$W), which are representative for the trades \cite{medeiros2016clouds}. Our analysis covers December to May of 2002 to 2020. Our satellite dataset combines data from NASA's Moderate Resolution Imaging Spectroradiometer (MODIS) aboard the Aqua satellite, with data from the Clouds and the Earth's Radiant Energy System (CERES) instrument. We compute organization metrics from MODIS' cloud masks. For each cloudy scene, we calculate two metrics - cloud fraction ($f_c$) and degree of organization ($I_{org}$). The preprocessing of MODIS' cloud masks follows \citeA{janssens2021cloud}:  Scenes with $>20$\% cirrus coverage are excluded, as are cloud fields with solar zenith angles $>45^{\circ}$. In contrast to \citeA{janssens2021cloud}, we use the full 10$^{\circ}\times 10^{\circ}$ domain. Following \citeA{schulz2021characterization}, we focus solely on shallow clouds by excluding scenes with cloud-top heights $z_t>4\,$km. After preprocessing, approximately 750 cloud fields remain for analysis.
CERES provides hourly top-of-the-atmosphere SW and LW radiative fluxes for all-sky and clear-sky conditions, as well as cloud optical depth (${\tau}_{c}$), cloud albedo ($A_c$), cloud-top height ($z_t$), liquid water path ($\mathcal{L}$), and cloud-droplet effective radius ($r_e$). We select CERES data around 13:30 local time, which corresponds to the overpass time of the Aqua satellite. Shortwave cloud radiative effect (SWCRE) and longwave cloud radiative effect (LWCRE) are calculated as the difference between the all-sky and clear-sky radiative fluxes at the top of the atmosphere. For each cloud scene, we calculate domain-mean values of cloud properties provided by CERES.

We amend our satellite analysis with the \textit{Cloud Botany} dataset \cite{jansson2023cloud}. This is a large ensemble (ca.~100~members) of high-resolution (100 m) large-eddy simulations (LES) of shallow cumulus clouds with a domain size of 150 km by 150 km. It was initialized with a variety of conditions derived from ERA5 reanalysis data \cite{hersbach2020era5} of trade cumuli that cover the climatological conditions of the area under consideration. We refer to the dataset paper, \citeA{jansson2023cloud}, for details. Our motivation for employing the \textit{Botany} simulations is twofold. Firstly, considering that liquid-water path $\mathcal{L}$ and effective radius $r_e$ might be underestimated in less organized conditions of broken cloud fields containing small clouds \cite{zhang2011assessment,seethala2010global,painemal2011assessment}, the simulations support that our results are physical. Secondly, the simulations provide data on cloud-base height ($z_b$) and cloud geometric thickness ($h$) so that we can investigate how they are correlated to organization. 
We use hourly data from hours 37-43 of the simulations (574 cloud fields in total). These times are chosen because they approximately align with the daily overpass times of the Aqua satellite to match the diurnal phase. To determine the geometric thickness $h$ of each cloudy column, we calculate the difference between the altitudes of the highest and lowest cloudy pixels where the liquid water specific humidity is larger than zero. Subsequently, for each cloud field, we compute the domain-averaged $h$. We further compute the mean size of cloud objects within each cloud field using: $L_c = (\sum_{1}^{n} A_i) / n$, where $A_i$ represents the area of each individual cloud object $i$, and $n$ corresponds to the total number of cloud objects within the field.
\section{Results \& Discussions}
\label{results}

\subsection{Cloud organization impacts CRE independent of $f_c$ variability}
\label{sec:cre-org}
\begin{figure}[t!]
    \centering
    \includegraphics[width=\linewidth]{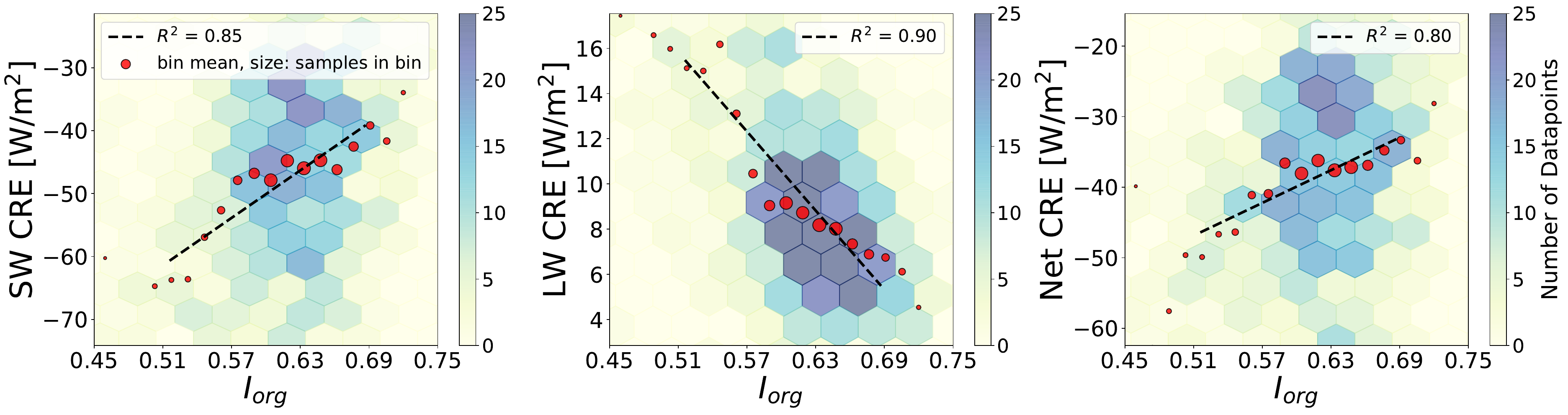}
    \put(-400,23){(a)}
    \put(-260,23){(b)}
    \put(-120,23){(c)}\\
    \includegraphics[width=\linewidth]{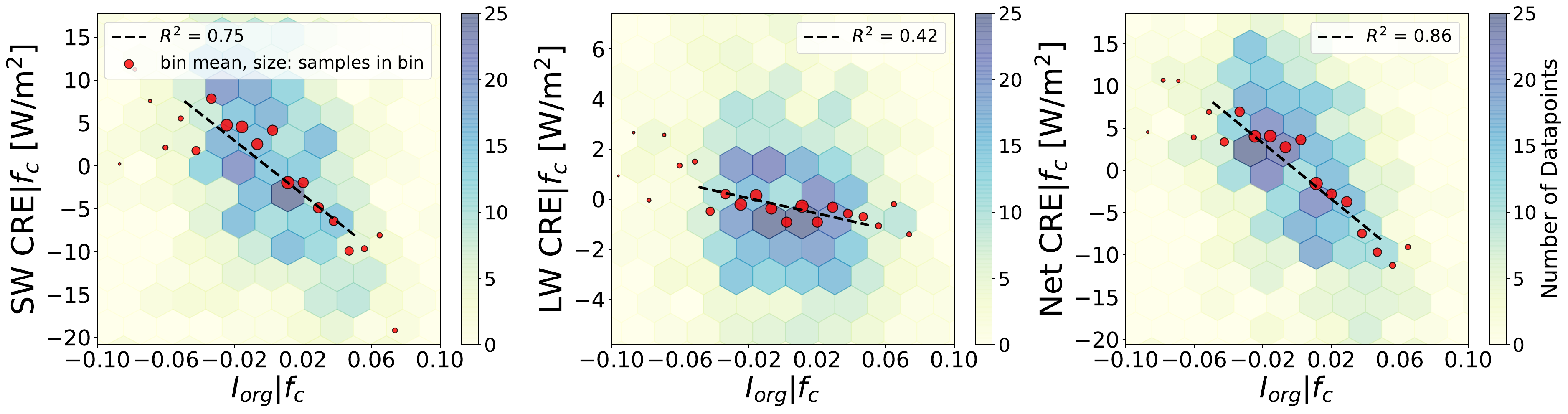}
    \put(-400,23){(d)}
    \put(-260,23){(e)}
    \put(-120,23){(f)}
    \caption{\textbf{Dependence of domain-mean CRE on organization.} In the first row (a-c), the relationships between $I_{org}$ and SWCRE (a), LWCRE (b), and net CRE (c) are illustrated. The second row (d-f) shows the same relationships but with removing the $f_c$ variability. The mean values of $I_{org}$ (for the $2^{nd}$ row, $I_{org} \vert f_c$) in each bin are denoted by red circles, with their size proportional to the number of points in the bin. The red dots are fitted with a dashed black line. Values below the $5^{th}$ and above the $95^{th}$ percentile of $I_{org}$ (for the $2^{nd}$ row, $I_{org} \vert f_c$) are excluded from the fit.}
    \label{fig:cre_org}
\end{figure}
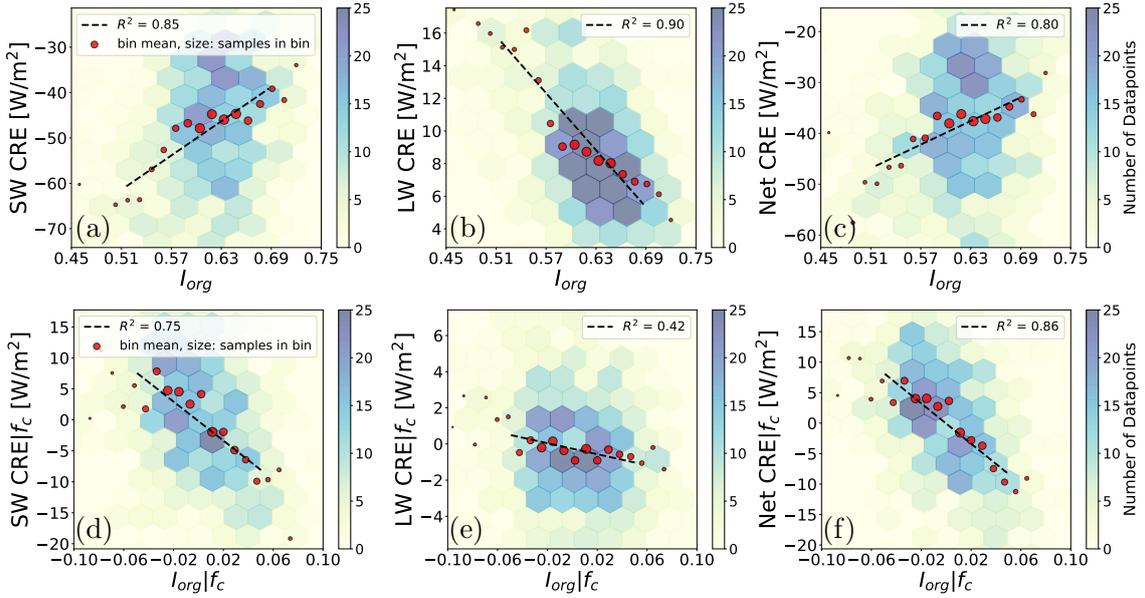

Figure \ref{fig:cre_org}(a-c) shows that as clouds become more organized (increasing $I_{org}$), they reflect less SW radiation towards space (smaller magnitude of SWCRE). In addition, increased organization also leads to a decrease in LWCRE. Overall, the warming effect induced by the SW component is partially compensated by the reduced LW warming. Consequently, enhanced organization of clouds results in diminished net cloud radiative cooling.

It is crucial to emphasize that the relationships illustrated in Fig. \ref{fig:cre_org}(a-c) are confounded by the variability of $f_c$, as $f_c$ is correlated with $I_{org}$, SWCRE, and LWCRE (Fig. S1). To eliminate the confounding effect of $f_c$, we employ the concept of partial correlation analysis \cite{baba2004partial}. For any given metric (e.g., $X$), we eliminate the variability associated with $f_c$ using a regression analysis,
\begin{equation}
X = c \times \text{$f_c$} + X \vert \text{$f_c$},
\label{eq: cf elimination}
\end{equation}
where $f_c$ serves as the regressor, $c$ represents the coefficient, and $X \vert \text{$f_c$}$ denotes the remaining variability in $X$ that cannot be explained by $f_c$. 

Fig. \ref{fig:cre_org}(d) shows that as $I_{org}$$\vert$$f_c$ increases, SWCRE$\vert f_c$ becomes more negative, i.e., as clouds cluster, they reflect more incoming SW radiation. The resulting SW cooling amounts to up to 20 W/m$^2$ . This indicates that the positive correlation observed in Fig. \ref{fig:cre_org}(a) is due to $I_{org}$ and SWCRE being negatively correlated to $f_c$ (Fig. S1, a, b). 
Similarly, after elimination of $f_c$ variability, the response of LWCRE to cloud organization is strongly reduced to about 1 W/m$^2$ (Fig. \ref{fig:cre_org}, e), indicating that the correlation between $I_{org}$ and LWCRE in Fig. \ref{fig:cre_org}(b) is almost solely due to their mutual correlation with $f_c$ (Fig. S1, c). The variability in LWCRE due to clustering is thus similar in magnitude to the LW radiative effect ($\approx$ 0.75 W/m$^2$) of the ``cloud twilight zone''  \cite{eytan2020longwave}. Ultimately, as Fig. \ref{fig:cre_org}(f) illustrates, the dependence of net CRE on $I_{org}$$\vert$$f_c$ arises almost exclusively from the dependence of the SW component on cloud organization. 

\subsection{Clustering and cloud optical thickness are positively correlated}
\label{sec:vert-org}
In the previous section, we eliminated the impact of $f_c$ on the $I_{org}$-SWCRE relationship. The remaining variability in SWCRE is caused by variations in albedo and could be modulated by a sensitivity of 3D radiative effects to cloud organization. Regarding the latter, \citeA{singer2021top} report that biases in top-of-the-atmosphere SWCRE from neglecting 3D radiative effect are negligible for both, unorganized and organized shallow cumulus cloud fields (their Fig. 7). This is especially expected for averages over large domains, as analyzed here.
Indeed, a bi-linear regression with $f_c$ and albedo as regressors can explain 94 \% of variability in SWCRE in our dataset (Fig. S2). This confirms that the impact of 3D radiative effects is negligible in these large cloud fields.

\begin{figure}[t!]
    \centering
    \includegraphics[width=\linewidth]{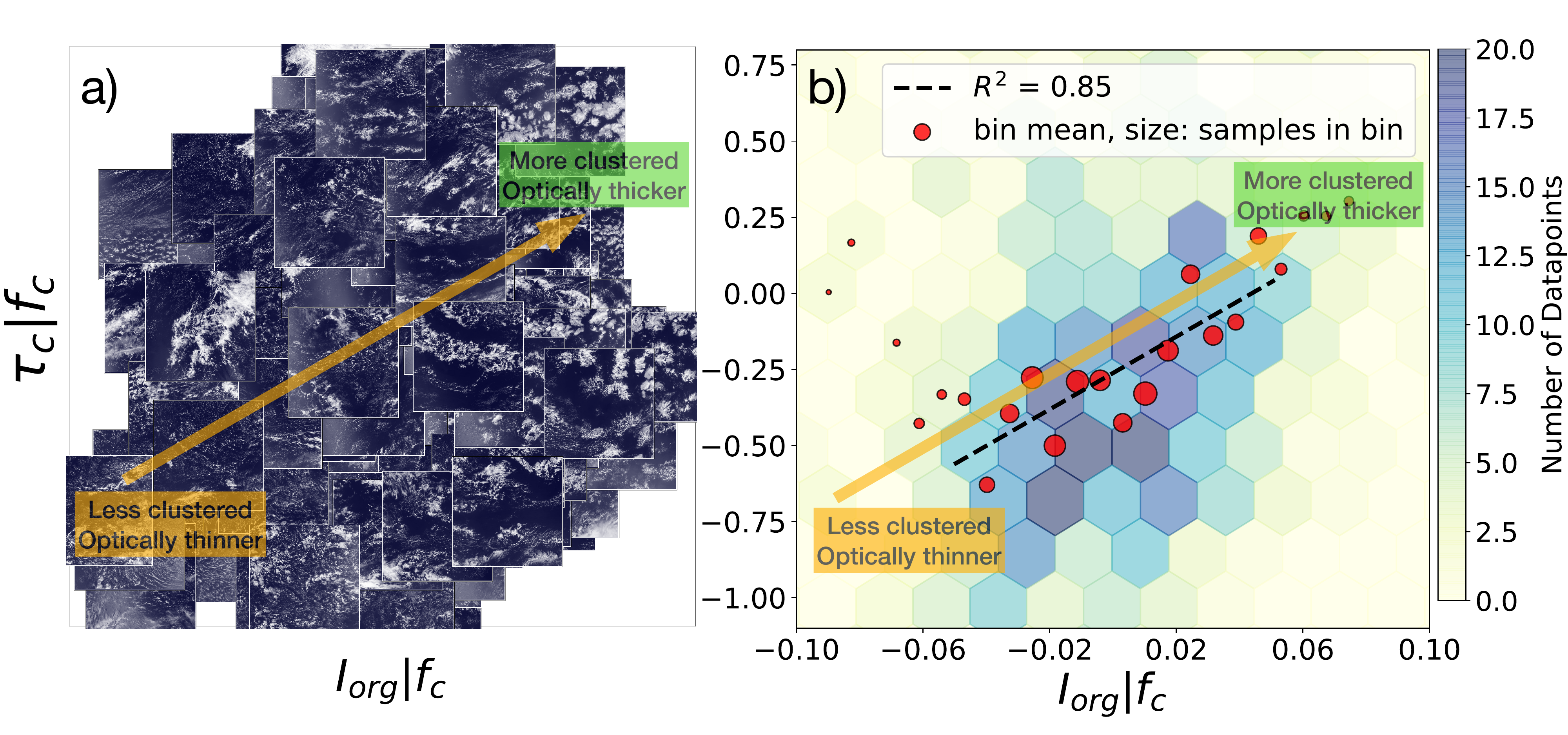}
    \caption{\textbf{Dependence of domain-mean cloud optical depth on organization.} The scatter plot of MODIS cloud features (a) and the 2-dimensional histogram (b) depict the $I_{org}$$\vert$$f_c$--${\tau}_c$$\vert$$f_c$ relationship. Specifically, for the scatter plot in (a), instead of displaying individual points, the entire cloud field is visualized to enhance pattern visualization. Clouds are represented in white, while the blue background represents the ocean color (MODIS true-color images). For plot (b), the mean values of $I_{org} \vert f_c$ in each bin are denoted by red circles, with their size proportional to the number of points in the bin. The red dots are fitted with a dashed black line. Values below the $5^{th}$ and above the $95^{th}$ percentile of $I_{org} \vert f_c$ are excluded from the fit.}
    \label{fig:cod_org|cf}
\end{figure}

Having 3D effects excluded, the remaining variability in SWCRE primarily corresponds to changes in albedo and equivalently cloud optical depth ${\tau}_c$. Figure~\ref{fig:cod_org|cf}(a) displays the variability of cloud patterns in a plane spanned by $\text{${\tau}_c$} \vert \text{$f_c$}$ and $\text{$I_{org}$} \vert \text{$f_c$}$. This figure shows a continuous range of patterns, in the terminology of \citeA{stevens2020sugar} ranging from \textit{Sugar} and \textit{Gravels} in the lower left corner to \textit{Fish} and \textit{Flowers} in the upper right corner.

On average, $\text{${\tau}_c$} \vert \text{$f_c$}$ increases with increasing $\text{$I_{org}$} \vert \text{$f_c$}$ (Fig. \ref{fig:cod_org|cf}, b). 
Despite the seemingly modest increase in $\text{${\tau}_c$} \vert \text{$f_c$}$, such changes can result in a significant increase in albedo of $0.03 \text{ to } 0.04$ (Figs. S3, S4), and the notable increase in SW reflection by 20 W/m$^2$ (Fig. \ref{fig:cre_org}, d). The $\text{$I_{org}$} \vert \text{$f_c$}$-$\text{${\tau}_c$} \vert \text{$f_c$}$ relationship indicates that horizontal cloud field organization, as measured by $I_{org}$, directly corresponds to its vertical organization, as captured by ${\tau}_c$: trade cumuli are optically thicker when they are more clustered.

\begin{figure}[t!]
    \centering
    \includegraphics[width=\linewidth]{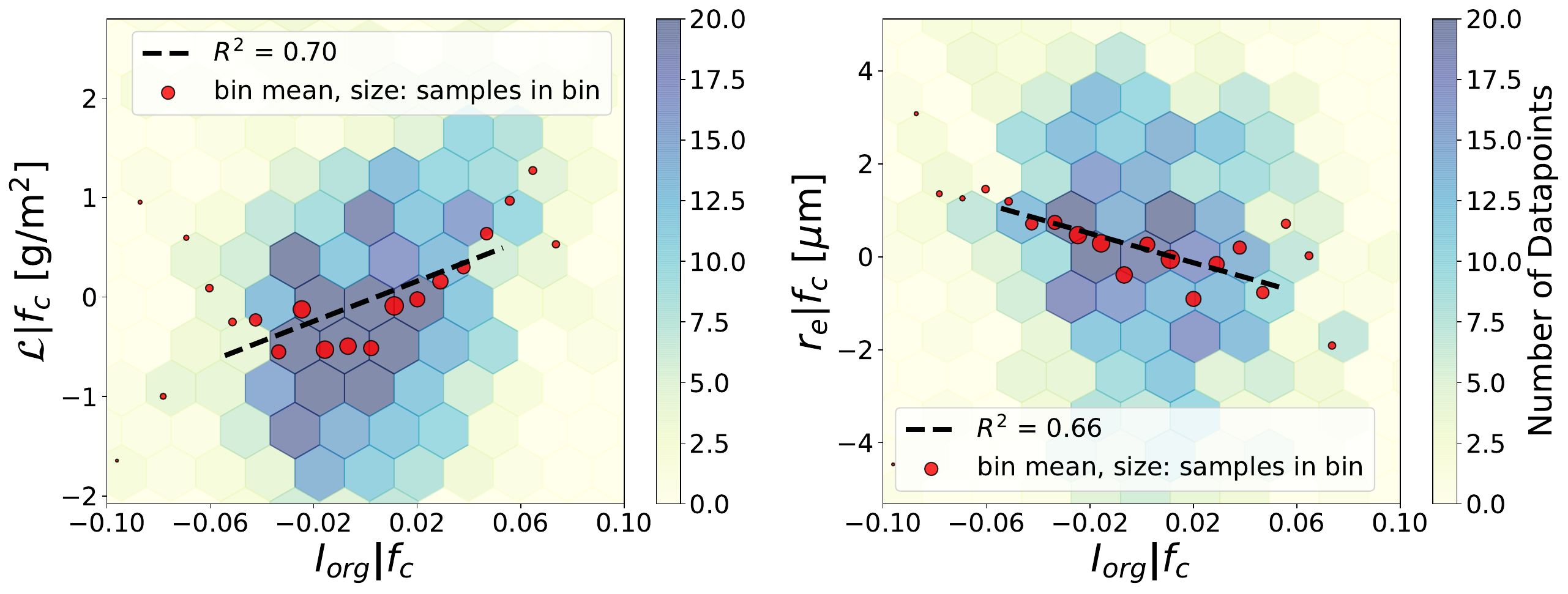}
    \put(-395,28){(a)}
    \put(-185,55){(b)}
    \caption{\textbf{Dependence of domain-mean liquid water path and effective radius on organization.} The figure shows the 2D histograms of the relationships between $I_{org} \vert f_c$ and $\mathcal{L} \vert f_c$ (a), and between  $I_{org} \vert f_c$ and $r_e \vert f_c$ (b). The mean values of $I_{org} \vert f_c$ in each bin are denoted by red circles, with their size proportional to the number of points in the bin. The red dots are fitted with a dashed black line. Values below the $5^{th}$ and above the $95^{th}$ percentile of $I_{org} \vert f_c$ are excluded from the fit.}
    \label{fig:lwp_reff_org|cf}
\end{figure}

As theoretically expected \cite{han1994near}, ${\tau}_c$ is proportional to ${\text{$\mathcal{L}$}}/{\text{$r_e$}}$ in our dataset (Fig. S5). Figure \ref{fig:lwp_reff_org|cf} shows that both, ${\text{$\mathcal{L}$}}$ and ${\text{$r_e$}}$, contribute to mediating the relationship between organization and optical depth. With the $f_c$ effect eliminated, there is a positive correlation between the degree of cloud clustering and the amount of liquid water present within the clouds (Fig.~\ref{fig:lwp_reff_org|cf}, a). Similarly, as the level of clustering increases, clouds tend to exhibit smaller radii $r_e$ (Fig.~\ref{fig:lwp_reff_org|cf}, b). A cloud field that is dominated by \textit{Flowers} shows smaller radii $r_e$ but higher liquid-water path $\mathcal{L}$ compared to the field dominated by \textit{Sugar} and \textit{Gravels}.

Our $\mathcal{L}$-organization relationship seems to be in contrast to \citeA{schulz2021characterization} who show that individual \textit{Gravel} clouds have higher liquid-water path $\mathcal{L}$ compared to individual \textit{Flowers}. To reconcile this with our results, we need to remind ourselves that the large cloud scenes analyzed here contain a mixture of different clouds. \citeA{stevens2020sugar} report that \textit{Gravel} clouds tend to coexist with \textit{Sugar}. For \textit{Flowers} such a coexistence is less pronounced. Instead, \textit{Flowers} feature anvils. Such shallow cumulus anvils have notable geometric thickness ($200$ m \cite{wood2018ultraclean} to 600 m \cite{dauhut2023flower}). This means that anvil cloudiness is optically thicker and more reflective than typical \textit{Sugar} (see also Fig. 7 in \citeA{stevens2020sugar}). When considering two cloud fields with identical $f_c$, it therefore seems reasonable that a \textit{Flower}-dominated field features a larger domain-averaged $\mathcal{L}$ as compared to a field dominated by \textit{Sugar} and \textit{Gravels}. 

Similarly, the relationship between $I_{org}$, $\mathcal{L}$ and $r_e$ might seem unexpected:
Based on adiabatic parcel lifting, we would expect $\mathcal{L}$ and $r_e$ to be positively correlated, while Fig. \ref{fig:lwp_reff_org|cf} suggests a negative correlation. In fact, $\mathcal{L}$ and $r_e$ are positively correlated in our dataset, while projecting their variability on organization ($I_{org} \vert f_c$) suggests a negative correlation: as degree of clustering increases, $\mathcal{L}$ increases but $r_e$ decreases (Fig. S6). Satellite snapshots of $r_e$ (Fig. S7) reveal that \textit{Flowers} have substantially smaller $r_e$ in their veils compared to their core updrafts. In contrast, \textit{Gravels} exhibit a more homogeneous $r_e$ with relatively larger values (compared to \textit{Flowers}) associated with strong updrafts at the edge of their cold pool fronts. These observations suggest that for \textit{Flowers}, the average $\mathcal{L}$ is primarily influenced by their cores, while the average $r_e$ is influenced by their veils. This is consistent with the fact that $\mathcal{L}$ is proportional to ${r_e}^6$, resulting in a more pronounced contrast between the core and veil in $\mathcal{L}$ compared to $r_e$. As a result, the average $\mathcal{L}$ of \textit{Flowers} is larger compared to that of \textit{Gravels}, while their average $r_e$ is smaller in comparison to \textit{Gravels}. 

It is interesting to contrast the small droplets in relatively thick anvils described here to the very large droplet sizes and optically thin veil clouds that have been reported in the context of the stratocumulus-to-cumulus transition \cite{wood2018ultraclean}. While \citeA{kuan2018deeper} report an increase in the corresponding ultra-clean conditions with boundary layer height, this relationship is unlikely to extend to deep trade cumulus \textit{Flowers}, which can be considered shallow mesoscale convective systems with complex outflow dynamics \cite{dauhut2023flower}. On the microphysical process level, ultra clean conditions have been associated with strong precipitation scavenging \cite{kuan2018ultraclean}, while \citeA{radtke2023spatial} discuss that the conversion efficiency to precipitation decreases with increasing organization in trade cumulus.

Overall, our discussion of the relationship between liquid-water path and effective radius to organization and the resulting effects on optical depth highlight that organized cloud fields cannot be conceptualized with a single, typical profile of cloudiness. Instead, the spatial variability requires considering two types of clouds as discussed for \textit{Gravel} and \textit{Sugar} versus \textit{Flower} cores and anvils. This duality resonates with the effective two-dimensionality of quantitative measures for mesoscale organization \cite{janssens2021cloud, shamekh2023implicit}.

\subsection{Mean cloud geometric thickness increases with clustering}
\label{sec:geo-org}

For an entraining lifting parcel, liquid-water path $\mathcal{L} \propto f_{ad}h^2$, where $f_{ad}$ represents the degree of adiabaticity and $h$ denotes geometric cloud thickness. To explain the observed $I_{org}$-$\mathcal{L}$ relationship, we therefore investigate the relationships of $I_{org}$ to $h$ and $f_{ad}$ in the \textit{Botany} simulations. Repeating the analysis from Sects.~\ref{sec:cre-org}, \ref{sec:vert-org} for the simulation data shows qualitative agreement and thus justifies this approach (Figs. S9(a-c), S10). Note that the discrepancy in the response of $r_e$ to cloud organization between simulations and satellite data (Fig. S9, d) is expected from the fixed cloud droplet number in the simulations but does not fundamentally affect our discussion of $\mathcal{L}$ here.

\begin{figure}[t!]
    \centering
    \includegraphics[width=\linewidth]{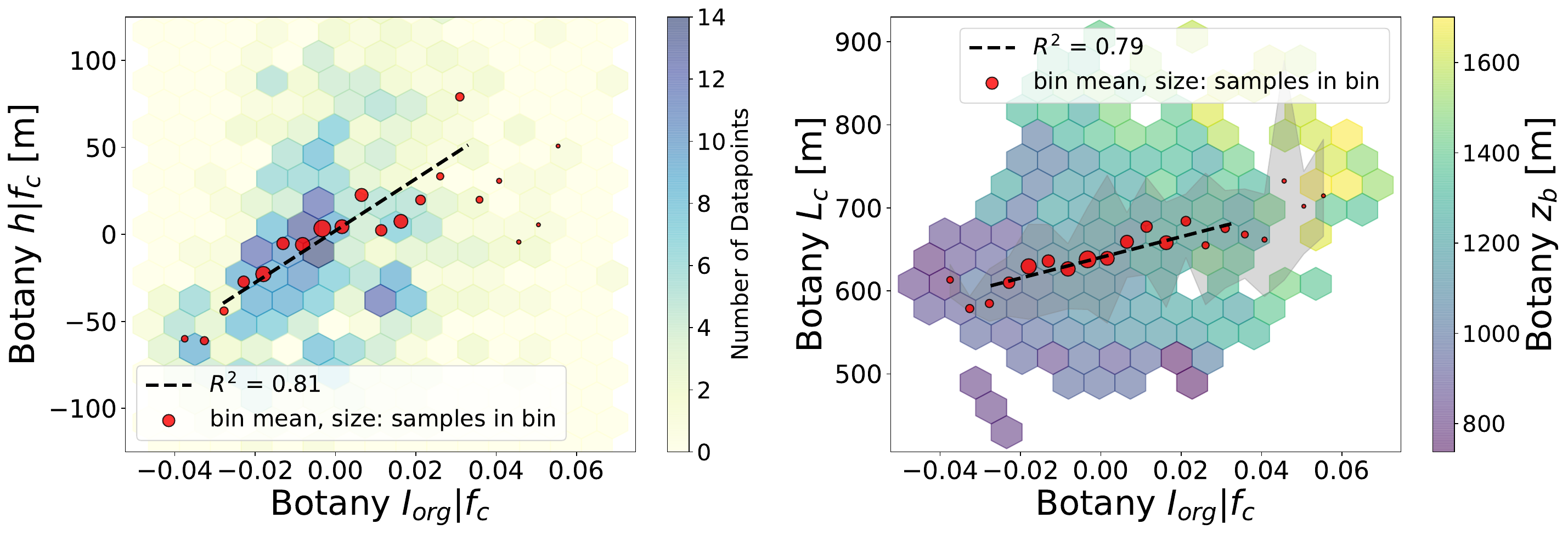}
    \put(-391,130){(a)}
    \put(-183,130){(b)}
    \caption{\textbf{Dependence of domain-mean geometric thickness, average size of cloud objects, and domain-mean cloud-base height on organization.} (a) The figure shows the 2D histogram of the relationship between $I_{org} \vert f_c$ and $h \vert f_c$. (b) The plot shows the relationship between $I_{org} \vert f_c$ and the mean-field cloud object size ($L_c$) with contour colors representing the values of domain-averaged cloud-base height ($z_b$). The gray shade indicates the inter-quartile range variability of $L_c$ in each bin of $I_{org} \vert f_c$. For both plots, the mean values of $I_{org} \vert f_c$ in each bin are denoted by red circles, with their size proportional to the number of points in the bin. The red dots are fitted with a dashed black line. For both plots, values below the $5^{th}$ and above the $95^{th}$ percentile of $I_{org} \vert f_c$ are excluded from the fit.}
    \label{fig:h_l_org|cf}
\end{figure}

Figure~\ref{fig:h_l_org|cf}(a) shows that the domain-averaged geometric thickness increases by more than 100 m as cloud fields become more clustered (increasing $I_{org} \vert f_c$). Additionally, the variability in $h$ has a significantly larger influence on the value of $\mathcal{L}$ than $f_{ad}$ (Fig. S11). Thus, our LES-based results indicate that the simulated increase in $\mathcal{L}$ due to enhanced clustering (Fig. S9, c) primarily stems from the geometric thickening of cloud fields.

Figure \ref{fig:h_l_org|cf}(b) further explores the relationship between horizontal and vertical cloud field properties and shows that the average size of cloud objects ($L_c$) increases with $I_{org} \vert f_c$. This positive correlation shows that cloud horizontal extent as quantified by $L_c$, is positively correlated to cloud vertical extent as quantified by $h$, consistent with the findings of \citeA{feingold2017analysis}. The figure moreover illustrates that an increase in $I_{org}$$\vert$$f_c$ corresponds to a rise in the domain-average cloud-base height ($z_b$). Note that $z_b$ is not the lifting condensation level; instead, it represents the lowest height of a cloudy pixel within each column. This means that a higher domain-mean $z_b$ is an indication of the presence of more anvils in the field. Overall, Fig. \ref{fig:h_l_org|cf}(b) demonstrates that enhanced clustering leads to a higher occurrence of larger cloud objects with elevated domain-mean $z_b$, indicating a higher degree of \textit{anvilness}.

\section{Conclusions \& Outlook}
\label{Conclusion}

\begin{figure}[t!]
    \centering
    \includegraphics[width=0.7\linewidth]{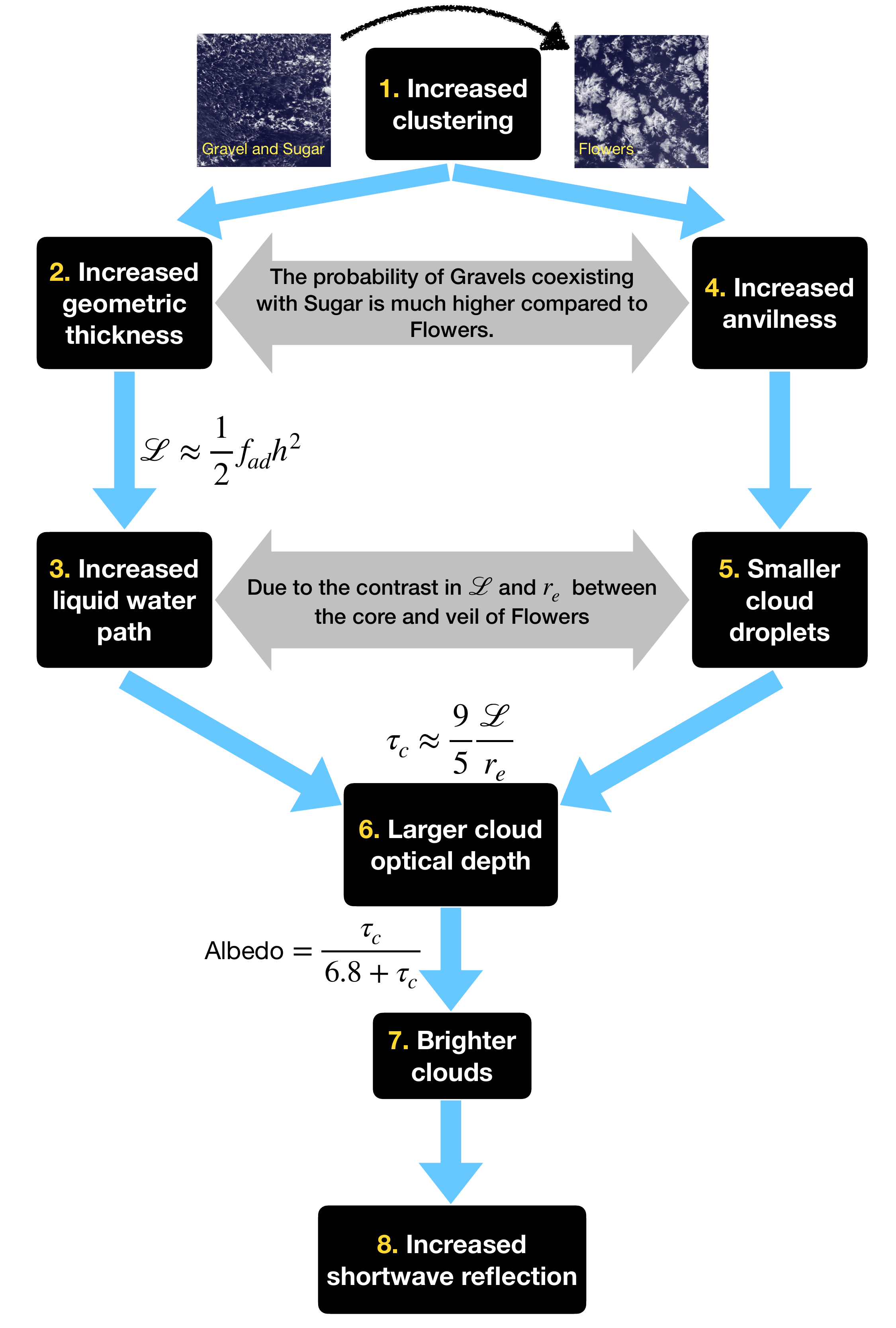}
    \caption{\textbf{Summary of results.} Comparing two cloud fields with identical cloud cover, the \textit{Flower}-dominated cloud field features a larger domain-averaged geometric thickness, a higher liquid water path, more frequent anvils with smaller cloud droplets, and consequently, brighter clouds and therefore larger SW reflection in comparison to the cloud field dominated by the presence of \textit{Sugar} and \textit{Gravels}. In the diagram, theory-based equations are provided for arrows where the underlying theory exists. The gray arrows illustrate apparent paradoxes which are discussed in section \ref{sec:vert-org}.}
    \label{fig:summary}
\end{figure}

We have explored the impact of shallow cumulus cloud field organization on cloud radiative effects, where confounding variability of $f_c$ was removed through partial correlation analysis (Eq. \ref{eq: cf elimination}). Based on satellite data, our analysis shows that an increased level of clustering ($I_{org} \vert f_c$) results in up to 20 W/m$^2$ higher SW reflection to space (Fig. \ref{fig:cre_org}, d, f). We observe that, irrespective of $f_c$ variations, more clustered cloud fields exhibit, on average, higher liquid water path (Fig. \ref{fig:lwp_reff_org|cf}, a), smaller cloud droplets (Fig. \ref{fig:lwp_reff_org|cf}, b), and consequently, greater optical thickness (Fig. \ref{fig:cod_org|cf}). A complementing ensemble of large-eddy simulations indicates that increased clustering leads to geometrically thicker cloud fields that feature increased anvilness (Fig. \ref{fig:h_l_org|cf}). Figure~\ref{fig:summary} summarizes these results. Collectively, they suggest that, eliminating the effect of $f_c$, the distribution of horizontal cloud sizes ultimately determines the vertical extent of clouds, subsequently influencing liquid-water path and cloud optical depth, and ultimately albedo and SWCRE.

What do our results mean in terms of the cloud feedback of trade cumulus? \citeA{myers2021observational} (their Supplementary Fig.~10) show that in addition to an increase in sea-surface temperature, which is not expected to trigger a notable response in trade cumulus cloudiness \cite{myers2021observational,cesana2021observational}, estimated inversion strength EIS is projected to moderately increase, and surface wind to slightly decrease. According to \citeA{bony2020sugar} such an increase in EIS would favor high-cloud-fraction \textit{Flowers} over \textit{Gravel} and \textit{Sugar} with lower cloud fractions. In contrast, the decreasing surface wind would favor \textit{Sugar}. While our results highlight the tight relationship between horizontal and vertical cloud organization, whether cloud fraction and optical depth interact positively or negatively in response to drivers of organization remains an open question. To address this interplay, we need to further explore how mesoscale processes \cite{janssens2023nonprecipitating,george2023widespread,vogel2021climatology} modulate cloud fraction, liquid-water path, effective radii, and anvilness.

\section*{Data Availability}
The cloud masks, provided by Aqua satellites, related to NASA's MODIS instrument, can be extracted from level-1 Atmosphere Archive $\&$ Distribution System Distributed Active Archive Center (\url{http:// dx.doi.org/10.5067/MODIS/MYD06_L2.061}). The data set related to CERES instrument is made available by Synoptic TOA and surface fluxes and clouds (SYN1deg - level 3) at \url{https://ceres.larc.nasa.gov/data/#syn1deg-level-3}. Prepossessing of cloud masks alongside calculation of organization metrics were done using the cloud metrics Gihub repository \cite{Denby_cloudmetrics} available at \url{https://github.com/cloudsci/cloudmetrics}. The Botany dataset was downloaded using the EUREC$^4$A intake catalog (\url{https://howto.eurec4a.eu/botany_dales.html}). The data was analyzed utilizing Python (used libraries: Numpy \cite{harris2020array}, Pandas \cite{mckinney-proc-scipy-2010}, Scipy \cite{2020SciPy-NMeth}, Matplotlib \cite{Hunter:2007}, and Seaborn \cite{Waskom2021}). ChatGPT (OpenAI: \url{https://openai.com/blog/chatgpt}) has been used for copy-editing during the preparation of the manuscript.

\section*{Acknowledgments}
PA is grateful to Lousie Nuijens and Stephan de Roode for fruitful discussions. FG and PA acknowledge support from The Branco Weiss Fellowship - Society in Science, administered by ETH Zurich. FG also acknowledges a NWO Veni grant. A. PS acknowledges support from the European Union's Horizon 2020 research and innovation program under grant agreement no. 820829 (CONSTRAIN project). TG acknowledges funding by the German Research Foundation (Deutsche Forschungsgemeinschaft, DFG; GZ QU 311/27-1) for project “CDNC4ACI”. GC acknowledges startup funds from Bar-Ilan University.

\bibliographystyle{apacite}
\bibliography{citations}

\newpage
\appendix
\section*{\centering \Large Supplementary Information for ``Shallow Cumulus Cloud Fields Are Optically Thicker When They Are More Clustered''}

\newpage
\noindent This file includes some additional figures for supporting the main text of our paper (First part). Additionally, the detailed calculation of adiabaticity is presented (Second part).

\subsection*{Supplementary Figures}

\setcounter{figure}{0} 
\renewcommand{\thefigure}{S\arabic{figure}}

\begin{figure}[p]
    \centering
    \includegraphics[width=\linewidth]{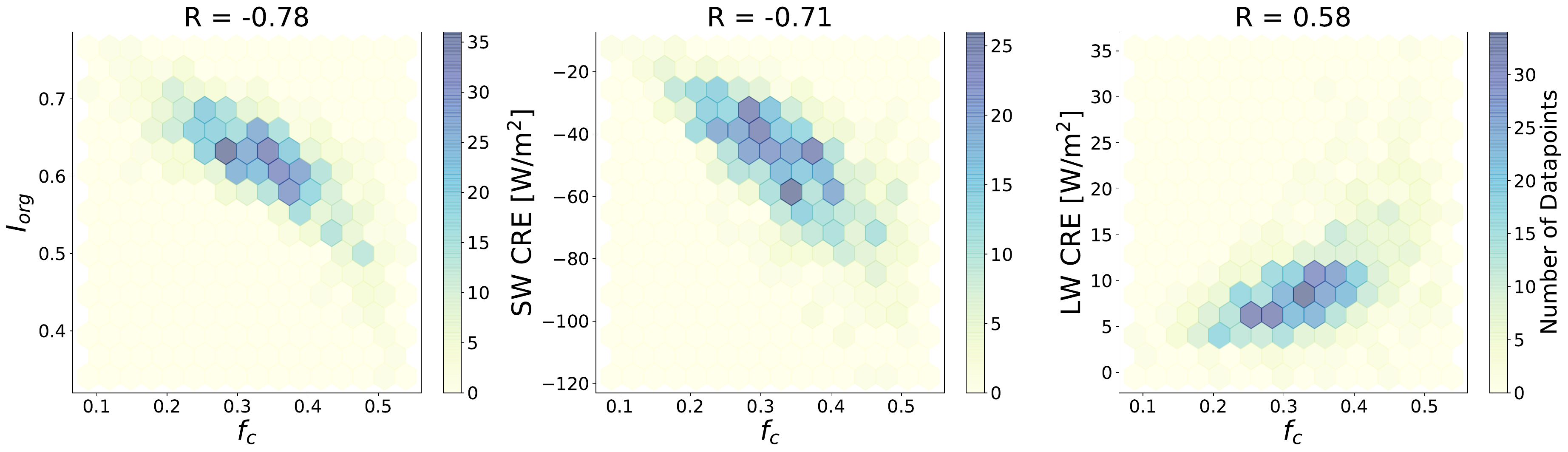}
    \put(-400,23){(a)}
    \put(-260,23){(b)}
    \put(-120,23){(c)}
    \caption{The 2D histograms of the relationships between $f_c$ and (a) $I_{org}$, (b) SWCRE, and (c) LWCRE with the reported Pearson's correlation (R).}
    \label{fig:cf_iorg_cre}
\end{figure}

\begin{figure}[p]
    \centering
    \includegraphics[width=\linewidth]{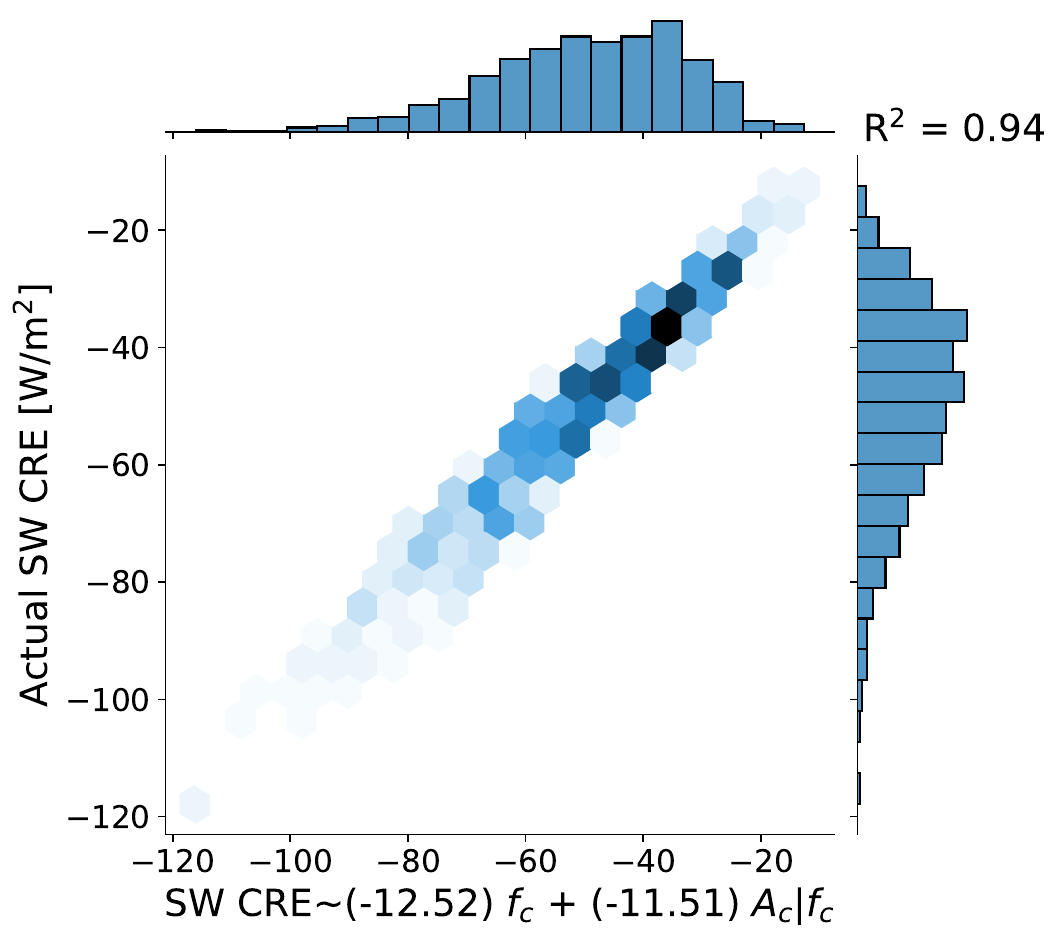}
    \caption{The result of the bi-linear regression analysis in which the target value is SWCRE and the regressors are $f_c$ and Ac$\vert$$f_c$ (Ac: cloud albedo observed by CERES). The reported coefficients are for the standardized $f_c$ and Ac$\vert$$f_c$.  This plots ensures that the derived $f_c$ from MODIS cloud masks in addition to the observed albedo by CERES can significantly capture the variability in SWCRE.}
    \label{fig:cf_alb_swcre}
\end{figure}

\begin{figure}[p]
    \centering
    \includegraphics[width=\linewidth]{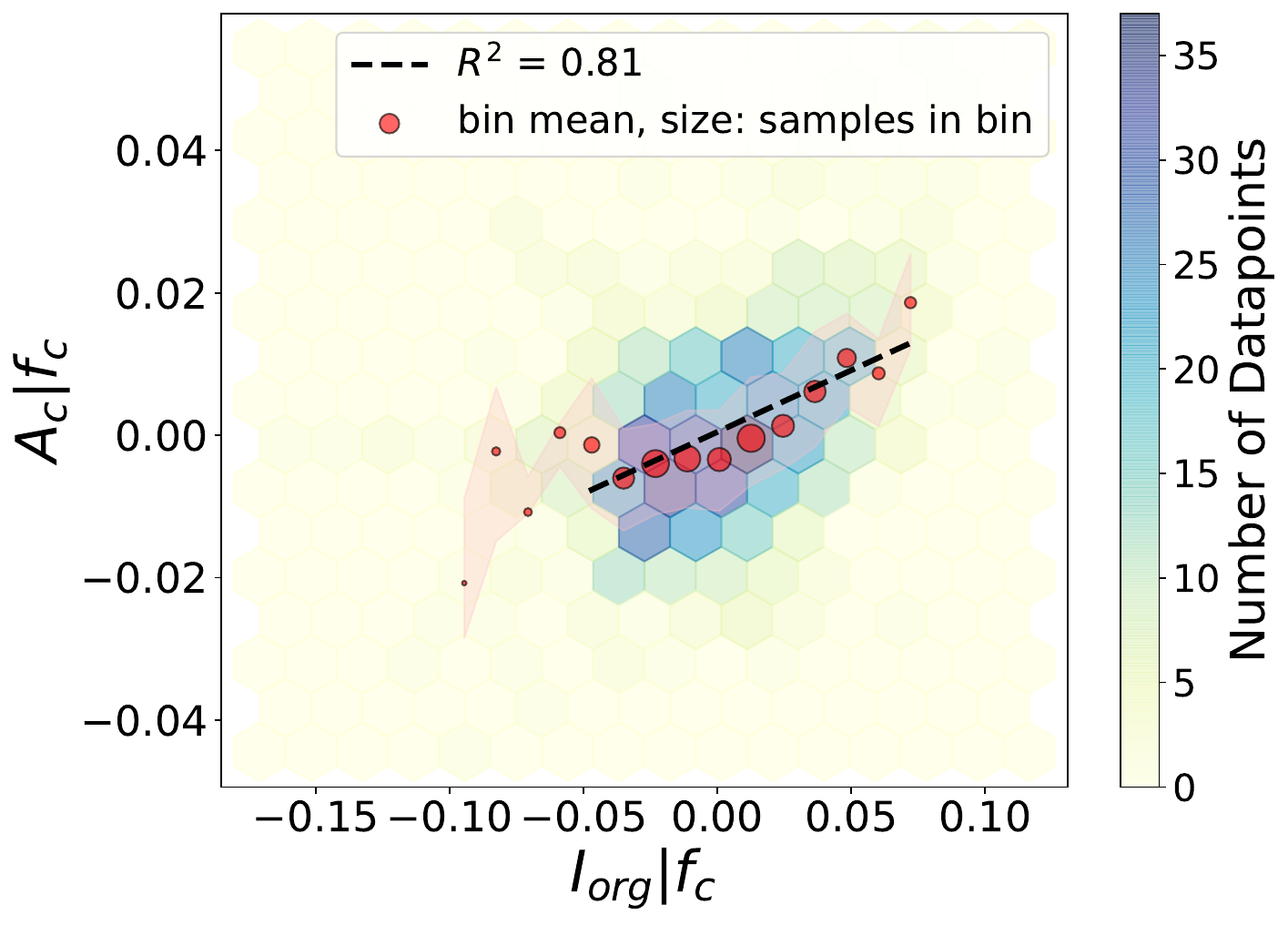}
    \caption{The relationship between $I_{org}$ and cloud albedo ($A_c$), having the effect of $f_c$ eliminated.}
    \label{fig:alb_org}
\end{figure}

\begin{figure}[p]
    \centering
    \includegraphics[width=\linewidth]{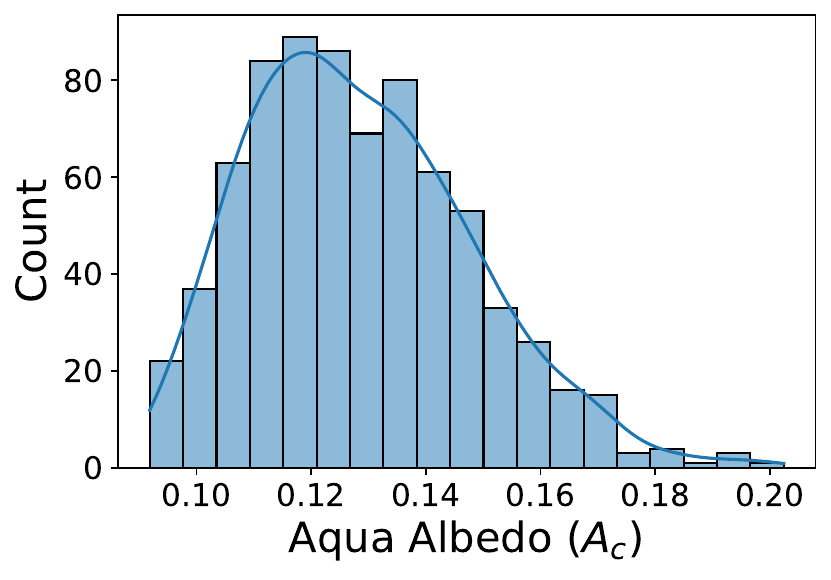}
    \caption{The histogram plot of cloud scene albedo ($A_c$) in the Aqua satellite dataset. This plot shows that the range of variability in cloud field albedo is about 0.1.}
    \label{fig:alb_hist}
\end{figure}

\begin{figure}[p]
    \centering
    \includegraphics[width=\linewidth]{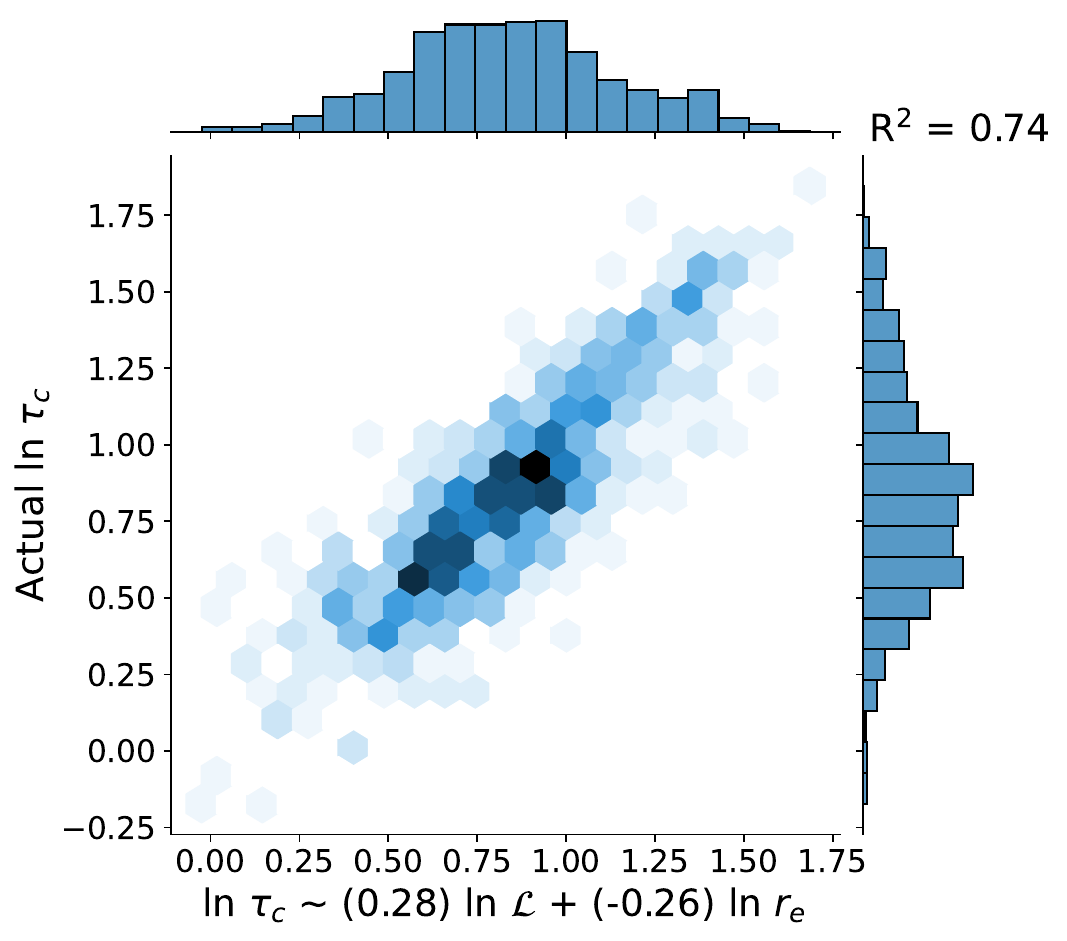}
    \caption{The result of the bi-linear regression analysis in which the target value is ln ${\tau}_c$ and the regressors are ln $\mathcal{L}$ and ln $r_e$. The reported coefficients are for the standardized ln $\mathcal{L}$ and ln $r_e$.  This plots ensures that ln $\mathcal{L}$ and ln $r_e$ can significantly capture the variability in ln ${\tau}_c$, in line with theory.}
    \label{fig:reff_lwp_cod}
\end{figure}

\begin{figure}[p]
    \centering
    \includegraphics[width=\linewidth]{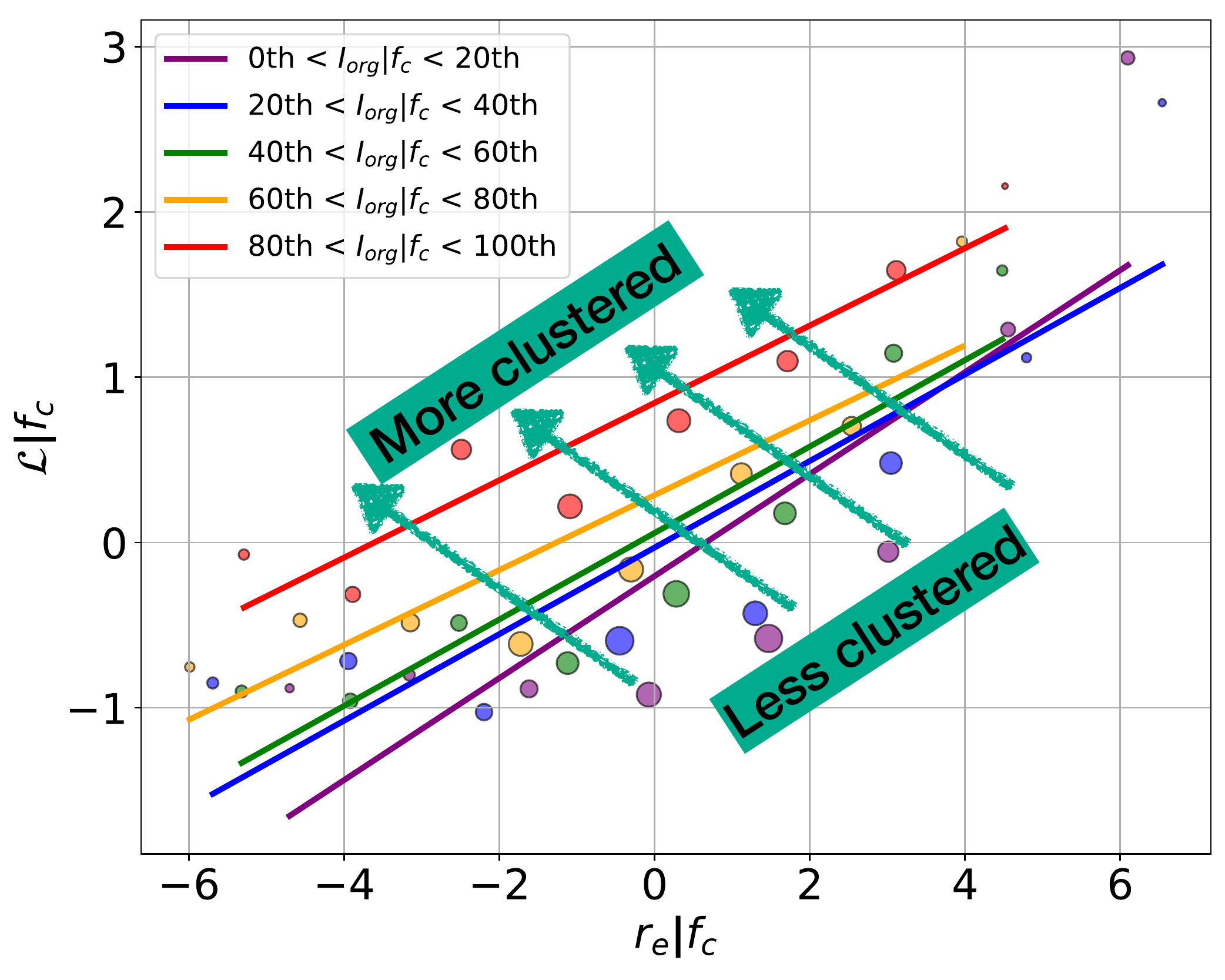}
    \caption{This figure highlights the positive correlation between $\mathcal{L}$ and $r_e$. However, when we examine the $\mathcal{L}$-$r_e$ relationship in relation to clustering, a clear pattern emerges: as clustering increases (more \textit{Flowers}), $\mathcal{L}$ increases while $r_e$ decreases. Each distinct color in the figure represents a specific class of organization. For instance, the purple color represents cloud fields where the value of $I_{org} \vert f_c$ falls within the range of the 0th to 10th percentile of $I_{org} \vert f_c$.}
    \label{fig:reff_lwp_iorg_paradox}
\end{figure}

\begin{figure}[p]
    \centering
    \includegraphics[width=\linewidth]{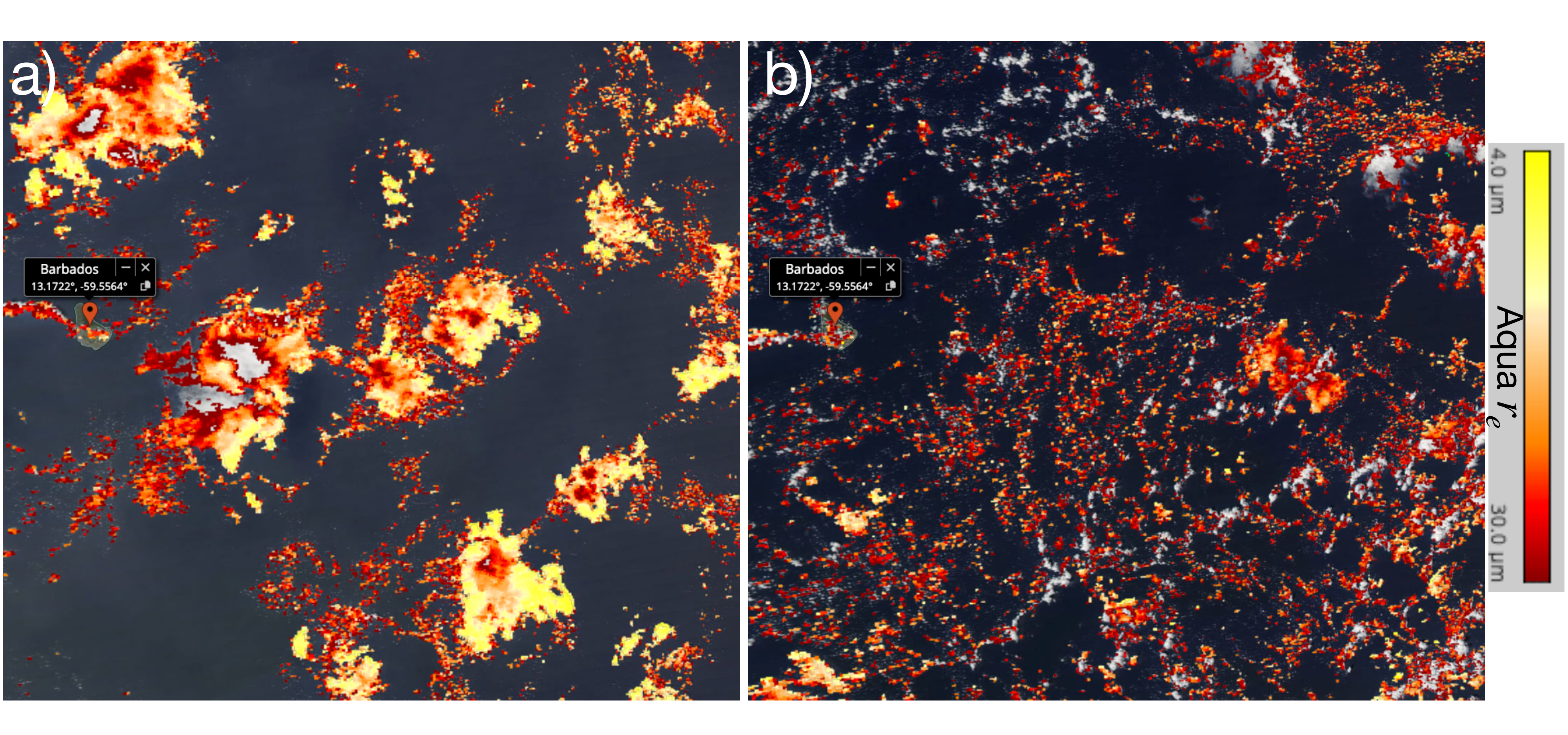}
    \caption{A Flowers (a) and a Gravel cloud field snapshots (from NASA Worldview) taken by Aqua satellites on 02/Feb/2020 and 17/Jan/2020, respectively. The cloud fields are colored by the value of $r_e$ which goes from 4 (yellow) to 30 (red) $\mu$m.}
    \label{fig:F_G_reff}
\end{figure}

\begin{figure}[p]
    \centering
    \includegraphics[width=\linewidth]{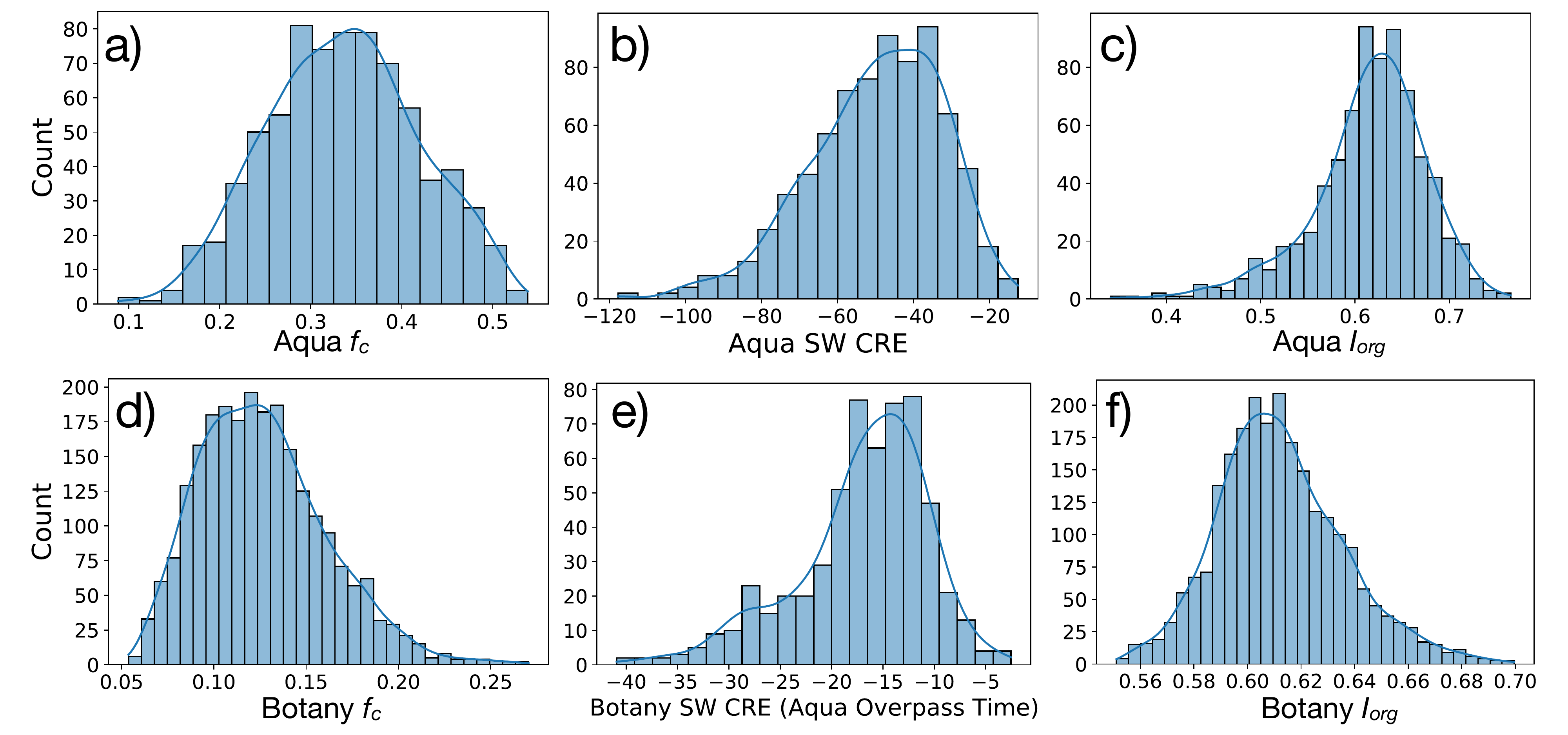}
    \caption{Histograms of $f_c$, SWCRE and $I_{org}$ for Aqua (a,b,c) and Botany (d,e,f) datasets, respectively. The values of SWCRE are in W/m$^2$. Note that the histograms of $f_c$ and $I_{org}$ are for the whole Botany dataset, while the Botany's SWCRE is only for hours 37-43 (day-time values).}
    \label{fig:hist plots}
\end{figure}

\begin{figure}[p]
    \centering
    \includegraphics[width=\linewidth]{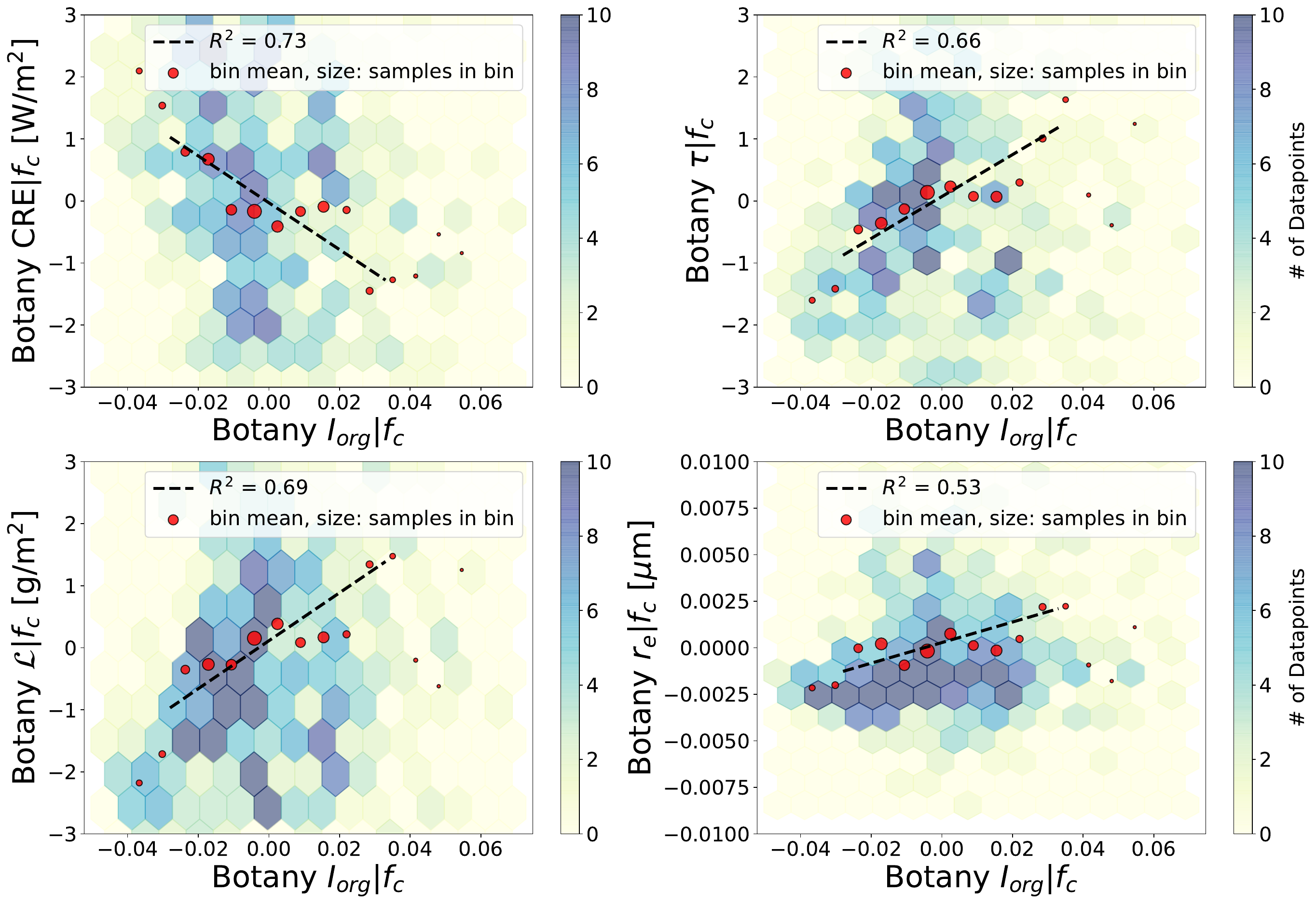}
    \put(-398,30){(c)}
    \put(-180,30){(d)}
    \put(-398,175){(a)}
    \put(-180,175){(b)}
    \caption{All results derived from the hourly analysis of the Cloud Botany dataset during the second day between the $37^{th}$ and $43^{rd}$ hours. Values below the $5^{th}$ and above the $95^{th}$ percentile of $I_{org} \vert f_c$ are excluded from the fitting.}
    \label{fig:botany plots}
\end{figure}

\begin{figure}[p]
    \centering
    \includegraphics[width=\linewidth]{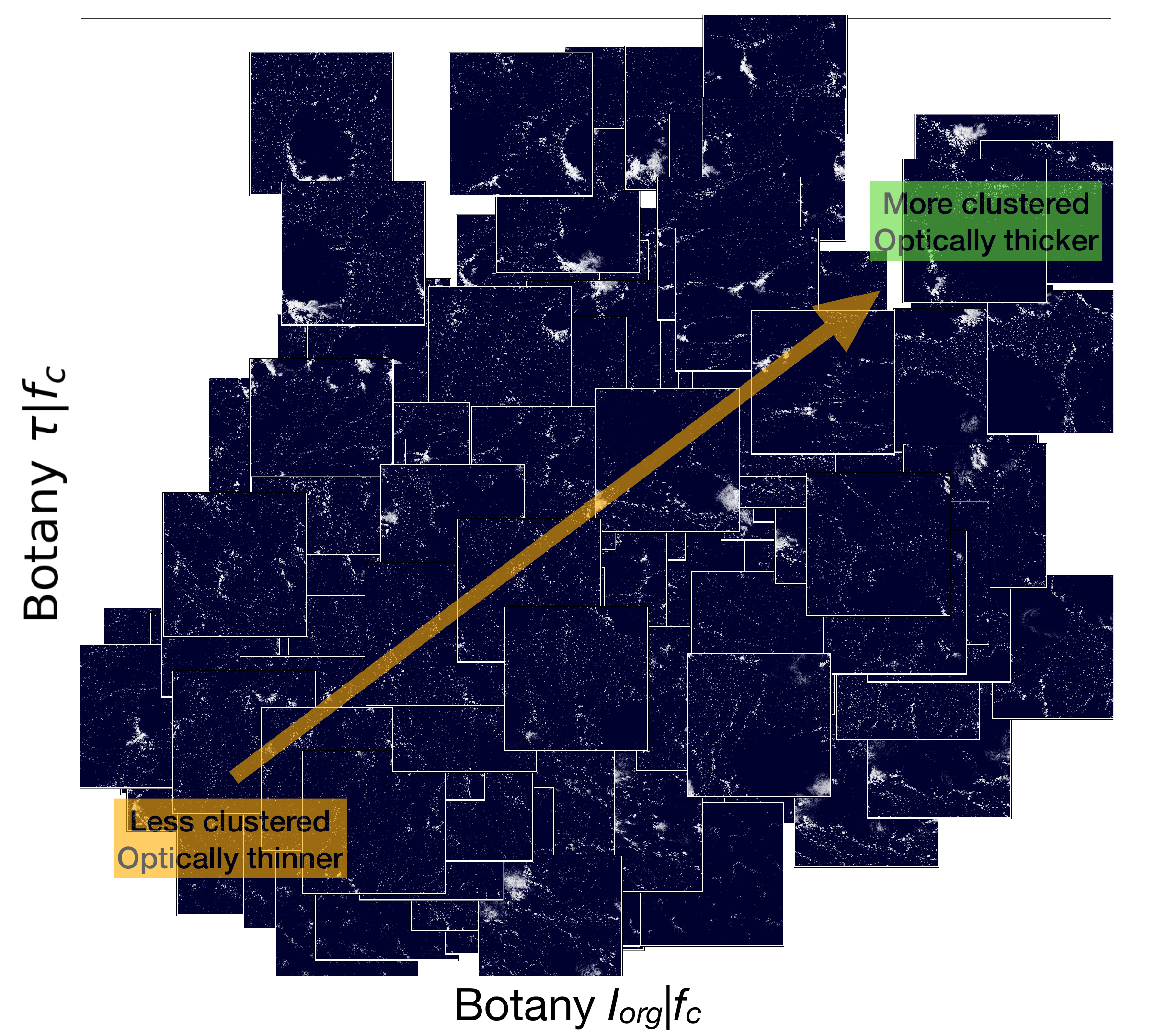}
    \caption{Similar to Fig. \ref{fig:cod_org|cf}(a) but for the Botany dataset during hours 37-43.}
    \label{fig:botany clouds}
\end{figure}

\begin{figure}[p]
    \centering
    \includegraphics[width=\linewidth]{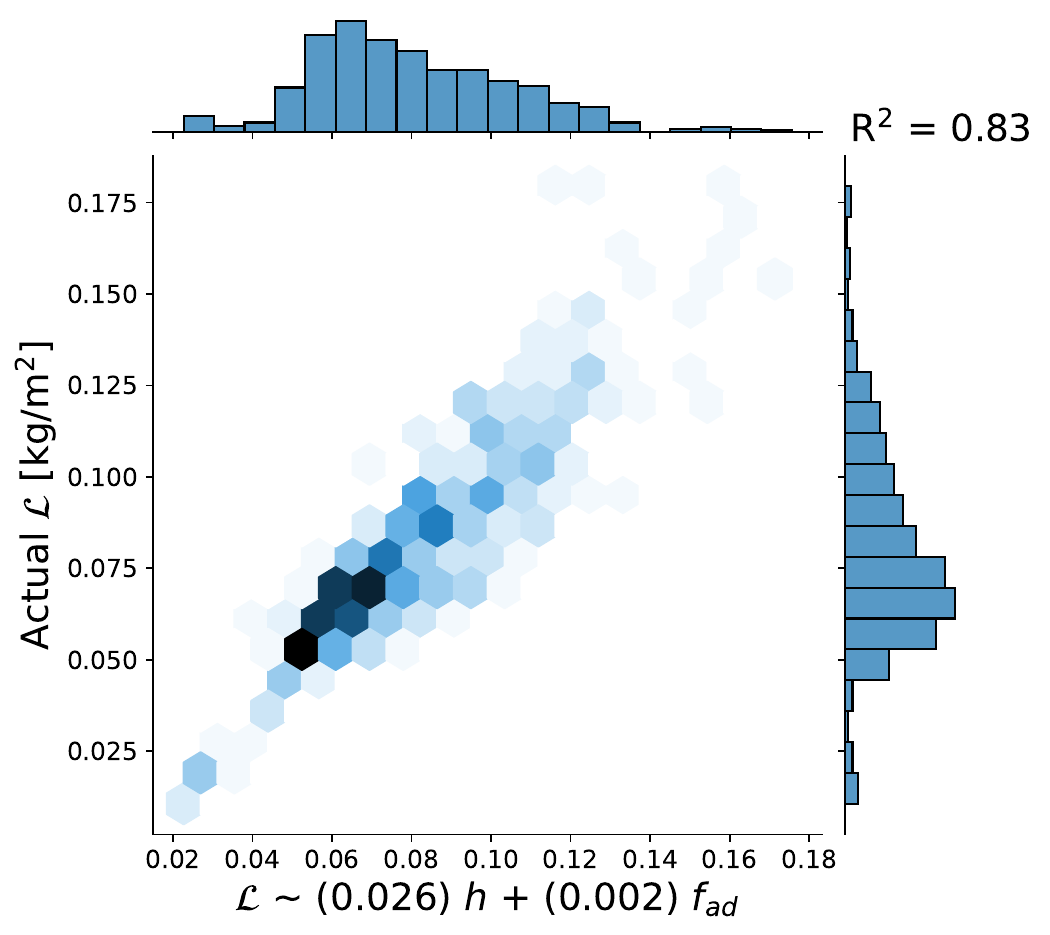}
    \caption{The result of the bi-linear regression analysis in which the target value is the in-cloud $\mathcal{L}$ and the regressors are $h$ and $f_{ad}$. The reported coefficients are for the standardized $h$ and $f_{ad}$.}
    \label{fig:h_fad_L}
\end{figure}

\newpage
\subsection*{Calculation of the degree of adiabaticity}
\label{apx:gamma}

The degree of adiabaticity is computed using the following equation:

\begin{equation}
f_{ad} = \dfrac{\int_{z_b}^{z_t} \rho(z) q_l(z) dz}{\int_{z_b}^{z_t} \rho(z) \Gamma_{ad}(z)zdz}
\end{equation}

where, $z$ represents the height, $\rho(z)$,  $q_l(z)$, and $\Gamma_{ad}(z)$ denotes density, liquid water specific humidity, and the moist adiabatic lapse rate of $q_l$ at each model height. To ensure a more precise estimation, $f_{ad}$ is computed only for columns where the associated $z_b$ is in proximity to the lifting condensation level ($<$850 m) and the cloud thickness exceeds 150 m.

The calculation of $\Gamma_{ad}$ is determined by the following equation \cite{schmeissner2015turbulent,eytan2021revisiting}:

\begin{equation}
    \Gamma_{ad} = \frac{c_p}{L_v} \left(\Gamma_m + \frac{g}{c_p}\right)
\end{equation}

Here, $c_p$ represents the specific heat constant, $g$ denotes the acceleration due to gravity, $L_v$ stands for the latent heat of vaporization. The moist adiabatic lapse rate for temperature, $\Gamma_m$, is computed as:

\begin{equation}
    \Gamma_m = \dfrac{g}{c_p} \left( \dfrac{1 + \dfrac{L_v}{R_dT}q_s}{1+\dfrac{\beta L_v}{c_p}q_s}\right)
\end{equation}

In the above equation, $R_d$ corresponds to the gas constant for dry air, $T$ represents temperature, $\beta$ is the constant in the Clausius-Clapeyron equation ($e_s = A\exp(\beta (T-T_0))$), and $q_s$ denotes the saturation specific humidity, which can be approximated as $\approx 0.622 \frac{{e_s}}{{P}}$ where $P$ represents pressure. 

For simplicity, we calculate $\Gamma_{ad}$ for the middle of the cloud layer (at $z' = \dfrac{z_b+z_t}{2}$) and assume that it linearly changes with height through the cloud layer. This assumption leads to the equation below:

\begin{equation}
f_{ad} = \dfrac{\int_{z_b}^{z_t} \rho(z) q_l(z) dz}{\Gamma_{ad}(z')\int_{z_b}^{z_t} \rho(z)zdz}
\end{equation}

where, a $f_{ad}$ value of 1 in each column indicates a fully adiabatic cloud, while values smaller than 1 and larger than 0 indicate non-adiabatic clouds. In the final step, after obtaining $f_{ad}$ for each column, we proceed to calculate the domain-mean $f_{ad}$ for the entire cloud field.

\end{document}